# *In Vitro* Vascularized Liver and Tumor Tissue Microenvironments on a Chip for Dynamic Determination of Nanoparticle Transport and Toxicity

Alican Ozkan[1], Neda Ghousifam[1], P. Jack Hoopes[2], Marissa Nichole Rylander[1,3,4]

[1]     Department of Mechanical Engineering, The University of Texas, Austin, TX, 78712, United States

[2]     Department of Biostatistics and Medicine, Dartmouth College, Lebanon, NH, 03755, United States

[3]     Department of Biomedical Engineering, The University of Texas, Austin, TX, 78712, United States

[4]     Institute of Computational Engineering and Sciences, The University of Texas, Austin, TX, 78712, United States

**Corresponding Author:** Alican Ozkan; **Address:** 204 E. Dean Keeton St. 78712, Austin, TX, USA;

**Tel:** +1-512-806-9062; Email: alicanozkan@utexas.edu; **ORCID iD:**0000-0002-3825-5899

**Running Title:** Vascularized Liver and Tumor on a Chip

# Abstract

This paper presents the development of a vascularized breast tumor and healthy or tumorigenic liver microenvironments-on-a-chip connected in series. This is the first description of a vascularized multi tissue-on-a-chip microenvironment for modeling cancerous breast and cancerous/healthy liver microenvironments, to allow for the study of dynamic and spatial transport of particles. This device enables the dynamic determination of vessel permeability, the measurement of drug and nanoparticle transport, and the assessment of the associated efficacy and toxicity to the liver. The platform is utilized to determine the effect of particle size on the spatiotemporal diffusion of particles through each microenvironment, both independently and in response to the circulation of particles in varying sequences of microenvironments. The results show that when breast cancer cells were cultured in the microenvironments they had a 2.62-fold higher vessel porosity relative to vessels within healthy liver microenvironments. Hence, the permeability of the tumor microenvironment increased by 2.35- and 2.77-fold compared to a healthy liver for small and large particles, respectively. The ECM accumulation rate of larger particles was 2.57-fold lower than smaller particles in a healthy liver. However, the accumulation rate was 5.57-fold greater in the breast tumor microenvironment. These results are in agreement with comparable *in vivo* studies. Ultimately, the platform could be utilized to determine the impact of the tissue or tumor microenvironment, or drug and nanoparticle properties, on transport, efficacy, selectivity, and toxicity in a dynamic, and high throughput manner for use in treatment optimization.

## Introduction

Cancer is a major public health problem worldwide and is one of the leading cause of death in the United States (Siegel *et al.* 2018). Systemically delivered chemotherapy employed in combination with resection or radiation is the predominant treatment option for cancer (Blanco *et al.*, 2015). However, the non-selective nature of chemotherapy can cause significant toxicity (particularly to the liver), thus resulting in short term or chronic failure of organs and tissues (Jiang *et al.*, 2017, King & Perry, 2001, Macdonald, 1998). While only 0.5-1% of injected chemotherapy accumulates in the tumor, nearly 6-7% is deposited in the liver, which results in hepatotoxicity according to *in vivo* studies (Caballero *et al.*, 2017, Gangloff *et al.*, 2005, King & Perry, 2001, NDong *et al.*, 2015, Wilhelm *et al.* 2016). The conjugation of chemotherapy drugs with nanoparticles increases the accumulation of drug inside the tumor, while it decreases uptake by the liver (NDong *et al.*, 2015, Petryk *et al.*, 2013). Despite the potential of nanoparticles to enhance drug delivery, nanoparticle optimization is currently limited because physiologically representative systems do not exist to enable dynamic and spatial assessment of the impact of size, surface properties, and targeting ligands on biodistribution, efficacy, and toxicity. Whether a chemotherapy drug is used alone or conjugated with nanoparticles, a more comprehensive understanding of the dynamic transport characteristics of the drug as a function of specific microenvironmental conditions would enable improved drug formulation and targeted delivery to maximize therapeutic benefit and minimize toxicity. Conventional drug testing is performed using in *in vitro* cell culture systems first, followed by assessment with animal models. Two-dimensional (2D) cell culture systems are often used as initial model systems, due to their simplicity for evaluating the efficacy and toxicity of drugs. However, these systems do not recapitulate the evolving tumor microenvironment, which hosts multi-cellular and cell-matrix interactions (Antoine *et al.*, 2015, Buchanan *et al.*, 2013). Consequently, 2D cell culture neglects dynamically changing biomechanical effects, including matrix stiffening due to desmoplasia, increased compressive force due to growth, and elevated interstitial fluid pressure and altered fluid flow due to abnormal vasculature (Antoine *et al.*, 2015, Buchanan *et al.*, 2013, Szot *et al.,* 2011). These biomechanical effects are poorly understood, and they may decrease nanoparticle efficacy by reducing their transport and uptake (Buchanan *et al.*, 2013). Alternatively, animal models provide physiological fidelity that enables researchers to study transport mechanisms and distribution of drugs (King & Perry, 2001, NDong *et al.*, 2015, Petryk *et al.*, 2013). These models are commonly utilized to determine the efficacy and toxicity of drug doses and to observe biodistribution (Clarke, 2009, Fernandes & Vanbever, 2009, Patterson *et al.*, 2011). Nevertheless, there are limitations associated with animal models. These include animal expense (which limits high-throughput drug optimization), significant variability and differential response to therapy, and the inability to isolate the impact of specific microenvironmental conditions on transport and tissue response (Engelman & Kerr, 2012, Hintze *et al*., 2014, Kang & Kim, 2016, J. B. Kim, 2005, Rongvaux *et al*., 2014, Seok *et al*., 2013). Furthermore, animals and humans have different physiology, and recent studies report poor correlation between *in vivo* results from animal experiments and drug efficacy and toxicity data from humans (Engelman & Kerr, 2012, Fernandes & Vanbever, 2009, Hintze *et al*., 2014, Kang & Kim, 2016, J. B. Kim, 2005, Rongvaux *et al*., 2014, Seok *et al*., 2013). Thus, there is a critical need for a chemotherapy test system that accurately represents the human microenvironment, facilitates high-throughput screening, and informs the optimization of drug uptake, efficacy, and toxicity.

Three-dimensional (3D) *in vitro* models provide physiological environmental conditions for tumor and tissue environments by influencing flow, vessel properties, and particle characteristics on drug transport and uptake (Cross *et al*., 2010, Fischbach *et al*., 2007, Fischbach & Mooney, 2007, Szot *et al.,* 2011). 3D *in vitro* models also enable dynamic and spatial assessment of drug/nanoparticle transport and cell response (Farokhzad *et al*., Ghousifam *et al.* 2015, 2005, Ng & Pun, 2008, Shin *et al*., 2013). Thus, clinically relevant *in vitro* microenvironments can be created to replicate human tissues both physiologically and pathologically, with the advantage of a more controlled environment to study cell toxicity mechanism, diseases, and drug screening both spatially and temporally (Song *et al*., 2012, Zervantonakis *et al*., 2012, Zheng *et al*., 2012). 3D *in vitro* breast tumor models have been utilized in biomedical research specifically for drug screening and transport studies. Collagen-based vascularized microfluidic breast tumor microenvironments have been created to generate chemoattractant gradients between multiple vessels, enabling characterization of cell mobility and extravasation (M. B. Chen *et al*., 2013, Jeon *et al*., 2015, Pavesi *et al*., 2016, Zervantonakis *et al*., 2012). Although these studies reported permeability coefficients of the tumor tissue and vessel in response to different treatments, vessels were not fully surrounded by extracellular matrix, which limited physiological response and estimation of transport spatially. To address these concerns with the clinical relevance of existing 3D models, we have created microfluidic breast tumor microenvironments in which tumor cells are cultured in collagen surrounding an endothelialized vessel in which physiological flow can be introduced (Buchanan *et al*., 2014, 2013, Michna *et al*., 2018). These platforms have been utilized to determine the influence of flow and tumor-endothelial crosstalk on vessel permeability and angiogenesis but they have not yet been applied to critical questions of chemotherapy drug and nanoparticle transport.

As described above, knowledge of the relative action of a drug on tumors compared to liver tissue is vital for the development of successful chemotherapeutic agents. Several groups have created tissue-on-a-chip systems consisting of multiple tissues to investigate the interplay between liver and tumors (Ma *et al*., 2012, Sung *et al*., 2013, 2010, Viravaidya *et al*., 2004, Wang *et al*., 2006). Novel microfluidic platforms were developed to estimate drug toxicity when different cell types were cultured as single monolayers on top of a polymethyl methacrylate surface connected with microchannels without the presence of physiologically representative extracellular matrix (ECM) (Sung *et al*., 2013, 2010, Viravaidya *et al*., 2004). Other studies have created polylactic acid (PLA) scaffold-based platforms connected with microchannels made of polymethyl methacrylate to study metabolization of chemotherapy drugs by the liver and for determination of tumor toxicity (Ma *et al*., 2012, Wang *et al*., 2006). However, the use of non-native extracellular matrix materials limits physiological cell-matrix interactions that significantly contribute to cell adhesion, proliferation, and representative cell response (Antoine *et al*., 2015). Although many of these platforms contain channels to simulate transport, these channels are not endothelialized, and the artificial boundaries between the channel and the surrounding cells limit the information that can be gained regarding particle transport (Buchanan *et al*., 2014, 2013). Kamm *et al*. developed unique collagen-based vascularized tumor microenvironments, but these platforms were not adapted to enable study of the interaction between liver and tumor sites (M. B. Chen *et al*., 2013, Jeon *et al*., 2015, Pavesi *et al*., 2016, Zervantonakis *et al*., 2012). Therefore, there is a critical need to develop a physiologically representative tissue-on-a-chip system that enables transport, toxicity, and efficacy for vascularized tumor and liver microenvironments to be assessed.

In this work, we have developed multiple novel tissue-on-a-chip platforms to simulate interactions between healthy or tumorigenic liver and breast tumor microenvironments. In doing so, we tested drug and nanoparticle development and assessed the dynamic transport of fluorescent nanoparticles in each compartment. We created multi tissue-on-a-chip platforms with healthy liver, liver cancer and breast cancer microenvironments. To mimic these microenvironments, cell lines of MDA-MB-231 for breast cancer, C3Asub28 for liver cancer, and THLE-3 for healthy liver were used. Microenvironments were created according to relevant mechanobiological factors. To demonstrate feasibility of microenvironments, cell viability was measured for three days, and native cell morphology was confirmed with SEM imaging and F-actin/DAPI staining. The fidelity of liver cells cultured in the microenvironment was demonstrated by detecting albumin expression and release in response to physiological shear stress. Dextran particles of size 3 and 70 kDa were perfused in the platform to replicate the hydrodynamic diameters of chemotherapy drugs and drugs conjugated with nanoparticles. The effect of different co-culture conditions on vessel permeability, ECM/vessel porosity, and accumulation of nanoparticles were quantified using intensity profiles in response to different interactions between breast tumor and liver microenvironments. Hence, we can simulate the conditions of drugs being metabolized (liver to breast tumor) and delivered directly (breast tumor to liver). Ultimately, the physiological multi tissue-on-a-chip platforms developed in this study enabled quantification of drug transport and distribution behavior, both spatially and temporally.

## Materials and Methods

**Human Cell Sources:** Human breast cancer cells (MDA-MB-231), healthy liver cells (THLE-3), carcinoma liver cells (C3Asub28), and telomerase immortalized microvascular endothelial (TIME) cells were used in this study. MDA-MB-231 cells (ATCC, VA, HTB-26) were cultured with Dulbecco's Modified Eagle's medium, nutrient mixture DMEM/F12(1:1) +L-Glutamine, +15mM HEPES (Invitrogen, CA) supplemented with 10% fetal bovine serum (FBS, Sigma Aldrich, MO), and 1% Penicillin/Streptomycin (P/S, Invitrogen, CA). TIME cells stably transduced with an mKate lentivirus were generously provided by the Wake Forest Institute of Regenerative Medicine, Winston-Salem, NC. These cells were cultured in Endothelial Basal Medium-2 (EBM-2, Lonza, MD) and supplemented with an Endothelial Growth Media-2 (EGM-2) SingleQuots Kit (Lonza, MD). Human liver carcinoma, C3Asub28 cells were generously provided by Dr Wei Li from the University of Texas at Austin. These cells were cultured in the same manner as the breast carcinoma cells. THLE-3 cells (ATCC, VA, CRL-11233) were cultured in BEGM Bullet Kit (Lonza, MD) with additional 5 ng/mL Epidermal Growth Factor (EGF, Invitrogen, CA), 70 ng/mL Phosphoethanolamine (Acros, Organics, Belgium), and 10% FBS in a pre-coated flask. All cells were incubated at 37°C and 95% atmospheric air/5% $CO_2$. Cell growth was monitored every day, and cells were detached when they were 70% confluent. All cell lines were used within the first eight passages.

**Tissue Properties and Preparation of Collagen:** Type I collagen was used as the primary extracellular matrix (ECM) component for each tissue microenvironment. The collagen preparation protocol is available in Supplement I. The mechanical property of ECM depends on collagen concentration, which also controls ECM porosity (Antoine *et al*., 2014). Since ECM stiffness directly affects cell-matrix interactions, such as cell adhesion/proliferation, and diffusivity of drugs into the tissue, it is critical to select an appropriate final collagen concentration to mimic the desired human tissue properties (Antoine *et al*., 2015). Yeh *et al.* reported stiffness of 3 kPa for hepatic tumor microenvironment (Yeh *et al*., 2002). Similarly, breast cancer tissue stiffness is reported as 4 kPa (Paszek *et al*., 2005). Chen *et al.* has shown the healthy human liver compression modulus varies between 0.59-1.73 kPa (Chen *et al*., 1996). Therefore, final collagen concentrations of 7 mg/mL were employed for liver and breast carcinomas, due to corresponding compression modulus of 3-6 kPa (Antoine *et al*., 2015, Buchanan *et al*., 2014, 2013, Szot *et al*., 2011). A collagen concentration of 4 mg/mL was selected to create the normal liver tissue, with corresponding compression modulus of 0.90-1.91 kPa (Antoine *et al*., 2015). 4 and 7 mg/ml collagen concentrations were also used for control studies.

**Device Design and Fabrication:** An aluminum mold, illustrated in Figure 1-I, was fabricated using micromilling technique. This fabrication method eliminates multistep fabrication processes and the necessity of expensive patterning reagents when compared to the conventional fabrication technique, photolithography (Michna *et al*., 2018). Well-mixed Polydimethylsiloxane (PDMS) with a curing agent ratio of 10:1 was poured inside the aluminum mold and baked for 1 hour at 75 °C (Figure 1-II). Solidified PDMS (which is the housing material) consisting of inlet and outlet channels, was peeled off from the mold and sterilized under UV for 1 hour with a glass slide before the bonding process. Then, the glass slide and PDMS were plasma-treated (Harrick Plasma, NY) and bonded to create the enclosure shown in Figure 1-III to surround the tissue microenvironment. To increase adhesion between collagen and PDMS housing, fabricated PDMS housing assembled with the glass slide was filled with sterile 1% Polyethylenimine (PEI, Sigma-Aldrich, MO) and incubated for 10 minutes. After aspirating PEI, channels were filled with 0.1% glutaraldehyde (Sigma-Aldrich, MO) and incubated for another 20 minutes. Glutaraldehyde was removed and the platform was washed twice with sterile DI $H_2O$. Collagen solution was neutralized to pH 7.4 with 1X DMEM, 10X DMEM and 1N NaOH and mixed with the intended cell line at a concentration of $1x10^6$ cells/mL. Collagen-cell mixture was injected into the platform to fill the enclosure. Final concentrations of collagen were selected as 4 and 7 mg/mL for healthy and tumorigenic tissues, respectively, to match human compression modulus of relevant tissue type as described in the previous section (Antoine *et al*., 2015). A needle was inserted inside the platform to form a hollow vessel before the polymerization of collagen as shown in Figure 1-IV. 3D illustration of fabrication schematic is provided in Figure S2-1 under Supplement II and further details of fabrication is reported in Supplement II. The needle sizes 22 and 27G (Jensen Global, CA) were used for tumor and healthy liver tissues, respectively, to provide the relevant physiological wall shear stress (WSS) in the tissues (Buchanan *et al*., 2014, 2013). Applied wall shear stress is a significant phenomenon to mimic human tissue as well as protein release by the cell lines (Buchanan *et al*., 2014). A clinical study performed by Korin *et al*. showed that human wall shear stress in vessels varies between 1-10 dyn/cm$^2$, and wall shear stress decreases down to 1 dyn/cm$^2$ for tumor microenvironments (Korin *et al*., 2015). However, wall shear stress higher than 4 dyn/cm$^2$ showed a reduction in albumin release according to an *in vivo* study conducted by Tanaka *et al*. (Tanaka *et al*., 2006). Therefore, needles at each given size were inserted for tumor and healthy liver microenvironments to provide 1 and 4 dyn/cm$^2$ wall shear stresses at the same flow rate. Details of numerical simulation and shear stress profile are provided in Supplement III. After the incubating the platform for 30 minutes at 37 °C and 5% $CO_2$, the collagen was polymerized and the presence of needles created a hollow vessel inside housings as illustrated in Figure 1-V, as previously published (Buchanan *et al*., 2013).

To create a fully functional aligned endothelium along each channel within each compartment, TIME cell suspension in media ($10x10^6$ cells/mL) was introduced in the channel (Figure 1-VI) and underwent flow preconditioning for 3 days (Buchanan *et al*., 2014, 2013). Within the first 36 hours, wall shear stress was maintained at 0.01 dyn/cm$^2$ and followed with a linear increase of wall shear stress to 0.1 dyn/cm$^2$ for 1 hour and maintained at this value for the next 36 hours. In the last 6 hours, wall shear stress was linearly increased to physiological value. For positive control samples, 20 ng/mL Tumor Necrosis Factor Alpha (TNF-$\alpha$, RnD Systems, MN) was perfused at 0.1 dyn/cm$^2$ for 24 hours after the preconditioning protocol. To provide flow into the microfluidic platform, 0.5" long 22G stainless steel needles were inserted through PDMS ports and partially into the collagen microchannels. Autoclaved Tygon silicone tubing (1/16" ID, Saint Global, France) was connected to the inlet needle and a bubble trap, which is connected to a syringe pump that controls the flow rate. The bubble trap eliminates the likelihood of washing out endothelial cells from the created vessel, with the effect of an introduced bubble in the platform channel. The outlet needle was similarly connected to silicon tubing, which collected the outlet media into a reservoir. Two chambers were connected using 22G pins and the same silicon tubing. Detailed images of the platform before and after assembly and preconditioning are shown in Figure 2.

**Cell Viability:** The viability was assessed in avascular platforms to measure growth kinetics of cells located in the ECM. The identical platform preparation protocol was followed as described in the previous section, without incorporation of endothelial cells. To maintain consistency, avascular platforms were cultured with endothelial cell culture media to maintain cell viability. Cell viability was measured using the CellTiter-Blue (Promega, WI) Assay, which gives direct correlation of cell metabolic activity to viable cells number. The viability was measured over the course of three days. Fluorescent intensity units were converted to cell concentration using the obtained calibration data presented in this work.

**Cell Morphology:** Cell morphology at days 0, 1, and 3 were determined as described previously (Szot *et al.*, 2011). Briefly, avascular platforms were fixed with 3.7% paraformaldehyde and permeabilized using 0.1% Triton X-100 (Sigma Aldrich, MO). Then, samples were blocked with 1% BSA (Santa Cruz Biotechnology Inc, CA) for 30 minutes at room temperature followed by an incubation step with rhodamine phalloidin (Invitrogen, CA), a high-affinity probe for F-actin. Samples were counterstained with DAPI (Vector Laboratories, CA) to visualize nuclei. Imaging was performed using Leica SP8 laser scanning confocal microscope. Another set of vascularized platforms were fixed to investigate cell morphology and ECM porosity using Scanning Electron Microscopy (SEM). SEM preparation protocol is provided in the Supplement IV.

**Albumin Expression and Release of Healthy Liver Cells:** Functional protein expression was assessed as is customary for other platforms (Buchanan *et al*., 2013, Ma *et al*., 2012, Sung *et al*., 2013, 2010). Rylander *et al.* has previously determined key angiogenic gene expression and protein release in a nearly identical vascularized breast tumor microenvironment (Buchanan *et al*., 2014, 2013). Since our objective was to assess the interaction between the vascularized tumor microenvironment and its vascularized liver counterpart, we assessed albumin expression by the liver. Albumin is the major protein expressed by liver cells and indicator of liver functionality (Braun *et al*., 1990, Esch *et al*., 2015). Albumin staining of THLE-3 cells seeded inside the vascularized microenvironment was carried out using FITC-tagged anti-human serum albumin antibody (Abcam, MA). Platforms were fixed, permeabilized, and blocked with the same method described in the cell morphology section. Diluted albumin antibody (1:200) was injected into the vessel of the platform and incubated for 1 hour at room temperature. Immunostained cells expressing albumin protein were imaged using a Leica TCS SP8 confocal microscope. Additionally, albumin release of THLE-3/TIME microenvironments was quantified using an Albumin enzyme-linked immunosorbent assay (ELISA) kit (Abcam, MA). Flow media samples were collected from the channel outlet at the end of each day of preconditioning. Additional samples were collected following the exposure of microenvironments to physiological wall shear stress for an hour after the preconditioning period. The measurements were carried out according to the manufacturer's protocol.

**Assessment of Transport Properties and Quantification:** Transport measurements of varying particle sizes in the multi tissue-on-a-chip microenvironment were conducted for two different scenarios: single microenvironment analysis and microenvironments connected in series to investigate the influence of their interactions and interdependent transport kinetics. When considering only a single microenvironment transport, particles were delivered through the vessel of the microenvironment of interest, and transport through the vessel and into the surrounding ECM was quantified subsequently and spatially. In these tests, six different platform configurations were used: acellular with no cells in the ECM and no cells in the vessel, TIME monoculture consisting of only endothelial cells lining the vessel without cells in the ECM with two different collagen concentrations (Control - Healthy and Tumorigenic), and then vascularized microenvironments denoted as cells in the ECM/cells in the vessel: MDA-MB-231/TIME, C3Asub28/TIME, and THLE-3/TIME microenvironments. In Control - (Tumorigenic) microenvironment, we set the collagen concentration to 7 mg/ml, vessel diameter is 711 $\mu$m and performed transport studies under 1 dyn/cm$^2$ wall shear stress. In Control - (Healthy) microenvironment, collagen concentration was set to 4 mg/ml, vessel diameter is 435 $\mu$m and transport studies were performed under 4 dyn/cm$^2$ wall shear stress. Microenvironments were connected in series to consider the influence of interactions between them. Particles were perfused through the first microenvironment's channel with associated diffusion into the corresponding ECM and back into the vessel, which resulted in transport to the next tissue compartment. We considered four different multi tissue-on-a-chip configurations: MDA-MB-231 to C3Asub28, MDA-MB-231 to THLE-3, THLE-3 to MDA-MB-231, and C3Asub28 to MDA-MB-231. Cases in which particles were introduced directly in the vessel corresponding to the breast tumor were used to simulate direct delivery to the breast tumor, where particles are not metabolized, and cases in which particles were first introduced into the vessel associated with the liver were used to simulate metabolization by the liver.

Passive transport of particles through blood vessels within the microenvironments depends on the permeability of each vessel endothelium and the porosity of the vessel and ECM of each tissue (Buchanan *et al*., 2014). Particle transport begins in the blood vessel, which is surrounded by endothelial cells and ECM. Endothelial integrity controls the barrier function and regulates transport of particles. According to *in vivo* studies, the gaps between endothelial cells are significantly higher in tumors vessels compared to healthy tissue vessels, and this is referred to as the enhanced permeability and retention (EPR) effect (Friedl *et al*., 2002, Jiang *et al*., 2017). Furthermore, this leakiness of the endothelium may also allow particles to diffuse back into the vessel from the ECM, which creates the vessel accumulation. Additionally, the ECM can act as a sink to trap particles, leading to accumulation within the tissue. Therefore, ECM and vessel porosity and permeability, which affect intravasation and extravasation of particles, need to be characterized to fully describe the expected transport of particles. The effect of porosities on diffusion is also stated by Darcy's Law given in Equation 1:

$$u = \kappa \nabla P / \zeta \mu \qquad (1)$$

where u is velocity in the porous domain, $\mu$ is viscosity, $\nabla P$ is the pressure gradient vector, and $\kappa$ is hydraulic permeability. This equation suggests porosity ($\zeta$) within the vessel and ECM determines the effectiveness of particle transport through ECMs. The velocity in the porous domain depends on porosities in each domain, which will consequently affect permeability and transport of the vessel and ECM. Therefore, we determined endothelial porosity using fluorescence microscopy images of mKate-tagged endothelial cells. ECM porosity is obtained by analyzing SEM images of ECMs as described in the previous section using ImageJ.

Selection of particle size is an important factor that controls the circulation time, tumor uptake, and ability of the particle to penetrate the tissue. Common chemotherapy drugs used for breast cancer treatment, such as doxorubicin, have hydrodynamic diameters in the range of 1.06-1.89 nm (Antoine *et al*., 2015, Blanco *et al*., 2015), which can also be calculated using the molecular weight and density of the drug. Although the hydrodynamic diameter of nanoparticle-conjugated chemotherapy drugs has great variability depending on the nanoparticle's type, size, and shape, it has been shown that common nanoparticle-conjugated chemotherapy drug size varies between 5-50 nm (Jiang *et al*., 2017). In this work, 3 and 70 kDa dextran particle sizes (Sigma-Aldrich, MO), with hydrodynamic diameter of 1.9 nm and 12.6 nm respectively, were selected to represent chemotherapy and nanoparticle-conjugated chemotherapy drugs, respectively, to demonstrate the EPR effect on the developed microenvironments (Antoine *et al*., 2015).

The effect of vessel and ECM porosities and particle size on transport were quantified using two methods: permeability coefficient and intensity profiles of the particles in the vessel and ECM. Fluorescent dextran particles suspended in serum-free (to prevent nanoparticle aggregation) endothelial basal media (EBM-2) to the final concentration of 10 µg/mL were perfused through the vascularized microenvironment for 2 hours with a flow rate of 260 µL/min, which yields physiologically representative shear stress in both microenvironments considered with appropriate vessel diameter. Images were taken every 3 minutes using a Leica SP8 Confocal Microscope. Obtained images were exported to Matlab® to quantify intensity readings at each time step. For the first method, the permeability coefficient was calculated using Equation 2 (Buchanan *et al*., 2014, Price & Tien, 2011)

$$P_d = \frac{I_2 - I_1}{I_1 - I_b} \frac{V}{S} \frac{1}{\Delta t} \qquad (2)$$

where $I_b$ is the background intensity, $I_1$ is the average initial intensity, $I_2$ is the average intensity after recovery, time interval $\Delta t$, and V/S is the vasculature volume to surface area ratio (Buchanan *et al*., 2014, Price & Tien, 2011). By definition, this parameter quantifies the ability of particles to penetrate from the microchannel to vessel wall, then to the ECM, and allows observation of the EPR effect. The last five consecutive data points from the two hours of flow were used to calculate permeability. For the second method, transport was quantified based on intensity profiles across the ECM boundaries when only one microenvironment was considered or when two microenvironments were connected in series with one another. Additionally, the same data were used to quantify the intensity change in the vessel and ECM to observe the rate of accumulation of different particle sizes in each compartment. Briefly, the rate of intensity change between each time step was calculated and averaged between t = 15 min and t = 120 min.

## Results and Discussion

In this work, we have developed the first vascularized multi tissue-on-a-chip microenvironments for modeling cancerous breast and cancerous/healthy liver microenvironments to allow for the study of dynamic and spatial transport of particles. Mechanical properties were tuned to mimic the native tissues modeled, and cell response, vessel permeability, and porosity of vessels and ECM were assessed. Ultimately, the transport kinetics and accumulation of varying sized fluorescent dextran particles, representative of chemotherapeutics and nanoparticle-conjugated chemotherapeutics within the tumor and liver microenvironments, were determined. The influence of particle delivery to specific tissue microenvironments to simulate direct tumor delivery, or metabolism of drugs prior to delivery to the tumor, was also investigated.

**Cell Morphology and Viability:** MDA-MB-231, THLE-3, and C3Asub28 cell lines were cultured in avascular collagen at concentrations mimicking each tissue's mechanical properties for three days, and cell morphology for each day was characterized. Figure 3 shows associated cell morphology using DAPI and F-Actin staining and SEM images. MDA-MB-231 cells developed an elongated, stellate morphology with disorganized nuclei, and invasive processes were observed by day three, as formerly reported for collagen-based *in vitro* platforms *in vivo* studies (Buchanan *et al*., 2013, Szot *et al.,* 2011, Xie *et al*., 2014). Comparably, THLE-3 cells exhibited an elongated morphology that is in agreement with isolated human hepatocytes (Pfeifer *et al*., 1993). Unlike the elongated healthy liver morphology, C3Asub28 liver cancer cells formed clusters, and the size of each cluster increased daily, as previously shown (Siveen *et al*., 2014, Sung & Shuler, 2009, Wang *et al*., 2006). SEM imaging was used to more clearly denote cell morphology and cellular interaction with the surrounding collagen matrix. In this assay, C3Asub28 cells were shown to possess a rounded shape, contrary to the epithelial THLE-3 morphology and the pleomorphic MDA-MB-231 cells with an elongated shape. Cell morphology was similar between immunostained images at day three and SEM images, which indicates that the SEM preparation did not affect cell and matrix properties. The noted morphological elongation of healthy liver cells and aggregation behavior of breast and liver cancer cells is due to cell-cell and cell-ECM interaction, as previously shown in collagen-based *in vitro* studies (Ma *et al*., 2012, Sung & Shuler, 2009, Szot *et al.*, 2011, Szot, Buchanan, Gatenholm, *et al*., 2011).

Figure 4 shows CellTiter-Blue viability results for each cell line over the course of three days. As shown, cells were viable over the time course of 3 days within the avascular collagen microenvironments. As anticipated, cells required some time to adhere before proliferating except for MDA-MB-231 cells, which proliferated significantly by 1.20-fold (p<0.05) on the first day. By the third day, C3Asub28 and MDA-MB-231 cells had proliferated by 1.23- and 1.34-fold (p<0.05), respectively. Although THLE-3 concentration did not change from day zero to day three, cells remained viable. These data confirm that the microenvironments support sustained cell viability, which is consistent with our previously published data with MDA-MB-231 in vascularized microenvironments (Buchanan *et al*., 2013).

**Albumin Expression and Release of Healthy Liver Cells:** The functionality of healthy liver cells was determined by detecting albumin expression and release. Albumin expression and release was measured for collagen-based vascularized THLE-3/TIME microenvironments for the first time in this study. Figure 5a shows anti-albumin immunostained THLE-3 cells in the collagen microenvironment. Cells exhibited elongated morphology, which is also shown in Figure 3. The albumin level presented in Figure 5b shows that the release increased significantly with time compared to day 1 ($p < 0.005$). While the number of cells did not change over the preconditioning period as illustrated in Figure 4, the increase of albumin expression can be explained by two main reasons. First, we observed in Figure 3 that cells exhibited a more elongated morphology with time, suggesting cells were becoming more established, yielding native genotypic and phenotypic behavior as reported by Szot *et al.* (Szot *et al.*, 2011). Second, increasing wall shear stress with each day is expected to promote greater cellular expression of albumin. Buchanan *et al.* previously showed that increasing wall shear stress promoted angiogenic gene protein expression of cells cultured in vascularized collagen platforms (Buchanan *et al.*, 2014). The albumin level of liver in normal human individuals is reported as 150-250 mg/kg/day (Braun *et al.*, 1990), and for human biopsy samples it is known that cell concentration is $0.65\text{-}1.85\text{x}10^8$ cells/g (Wilson *et al.*, 2003). Using the cell concentration and albumin level stated in these studies, cell-wise human albumin release was calculated as 0.81-3.85 pg/cell/day. Our measured albumin in response to physiological flow is $3.64\pm0.19$ pg/cell/day, which is within the range of published values. This verifies the functionality and fidelity of this developed vascularized healthy liver microenvironment under the given flow conditions.

**Porosity of Vasculature:** After embedding cells in collagen and successfully preconditioning endothelialized channels for 72 hours, the effect of different co-culture conditions on the vessel confluence was studied and shown in Figure 6. The first three cases involved creation of platforms with only a functional endothelium and no cells within the ECM, referred to as two TIME monocultures (Control –, Healthy and Tumorigenic) alone or in the presence of TNF-*α* being perfused in the vessel (Control +) to dilate the vessel pores for comparison. The last three conditions incorporated different cell types within the collagen ECM in addition to the TIME culture: C3Asub28/TIME, MDA-MB-231/TIME, and THLE-3/TIME microenvironments. A confluent endothelial lumen in which red fluorescence of mKate is shown with minimal dark gaps between cells is apparent for the Control – vessels. The vascularized endothelium co-cultured with THLE-3 shows a comparable endothelial confluency to healthy and tumorigenic Control – studies. Moreover, we observe that artificial modulation of the vessel with TNF-*α* treatment, (Control +), caused vessel permeabilization with significant pore openings compared to Control – (Tumorigenic). On the other hand, the tumor vessels (C3Asub28/TIME and MDA-MB-231/TIME) exhibit a patchy and leaky endothelium with perivascular detachment and non-uniform gaps unlike the uniform, dilated openings of Control +. This strengthens the idea that the cross-talk between cancer and endothelial cells cause a leaky porous domain, leading to the well-known EPR effect, also demonstrated in vascularized tumor microenvironments (Buchanan *et al*., 2014).

Vessel porosity of varying vascularized tissue microenvironments is reported for the first time in this study and presented in Figure 7a. Based on the measured results, inclusion of breast and liver tumor cell lines in the platform increased vessel porosity by 2.64- ($p<0.001$) and 3.62-fold ($p<0.001$), respectively, compared to the Control – (Tumorigenic). This is an evident phenomena that the cross-talk and signaling between cancer and endothelial cells and release of TNF-$\alpha$ resulted in detachment of endothelial cells that created large gaps around the vessel surface, as discussed in previous *in vitro* studies (Buchanan *et al*., 2014, 2013, Khatib *et al*., 2005, Zervantonakis *et al*., 2012). Our work also agrees with a prior 3D *in vitro* study, which showed that TNF-$\alpha$ can promote cancer cell transendothelial migration and invasion (Zervantonakis *et al*., 2012). We observed that co-culture with healthy liver cells (THLE-3) did not affect vessel porosity significantly compared to the Control – (Healthy). Therefore, the changes observed for endothelial integrity in cancer microenvironments compared to Control – are most likely due to signals provided by the cancer cells (Buchanan *et al*., 2013, Zervantonakis *et al*., 2012). One other reason for the observed patchy endothelial structure could be due to the heterogeneous distribution of cell clumps reported in Figure 3 liver and breast cancer cells located in the ECM. Cell aggregation leads to non-uniform release of expressed proteins across the ECM and vessel, resulting in cellular invasion into the endothelial layer over time, as described by previous collagen-based vascularized microenvironments (M. B. Chen *et al*., 2013, Jeon *et al*., 2015, Pavesi *et al*., 2016). The amount of released protein perfusing through the leakier endothelial layer of the liver cancer microenvironment is anticipated to be greater than that in the breast cancer microenvironment, due to 1.37-fold higher porosity ($p<0.05$) compared to breast cancer.

**ECM Porosity:** Following transport through the endothelium, nanoparticles or drugs must navigate the ECM to reach the tumor cell. Therefore, we also characterized the ECM structural properties of our microenvironments. Ramanujan *et al*. showed that fiber alignment is indirectly proportional to diffusive transport of particles (Ramanujan *et al*., 2002). SEM images exhibited in Figure 7b show fiber alignment is induced by shear stress during preconditioning, which is different than randomly oriented static ECM images (Szot *et al*., 2011). Quantified ECM porosity results presented in Figure 7c show that ECM porosity in MDA-MB-231 microenvironments did not change significantly, but C3Asub28 microenvironments increased by 1.14-fold ($p<0.05$) relative to Control – (Tumorigenic). Schedin *et al*. also previously indicated that *in vitro* mechanosignaling events carried out by cancer cells can alter ECM stiffness (Schedin & Keely, 2011) and consequently porosity, which has also been demonstrated in our study. As the type of cell line embedded in the collagen affects ECM porosity, drug transport behavior from the flow through the endothelium into the ECM is also expected to be altered correspondingly. Furthermore, THLE-3 microenvironment ECM porosity did not change significantly compared to Control - (Healthy). Control - (Healthy) ECM porosity is 1.29-fold (p<0.005) higher compared to Control - (Tumorigenic). Similar high porosity in lower collagen concentration was also reported in literature studies (Ramanujan *et al.* 2002, Wong *et al.* 2011).

**Vessel Permeability of Microenvironments:** Vessel permeability is an indicator of the leakiness of the endothelial lumen for each given condition as a function of particle size. Permeability was assessed using Equation 2 for two different dextran particle sizes (3 and 70 kDa) in 5 different types of microenvironments: acellular (no endothelial cells lining the vessel and no cells in the ECM), TIME monoculture (endothelialized vessel with no cells in the ECM), C3Asub28/TIME, MDA-MB-231/TIME, and THLE-3/TIME. Permeability findings were presented in Figure 7d. Our results show that permeability is higher for acellular (cell-free) microenvironments for both particle sizes due to the lack of an endothelial barrier. Furthermore, despite having higher ECM porosity, permeability decreased significantly when THLE-3 cells were cultured in the ECM compared to Control – samples. This is likely due to the THLE-3 microenvironment having higher shear stress, which gives less time for particles to diffuse through the ECM (Buchanan *et al*., 2014). Therefore, we did observe a notable permeability decrease with respect to Control – (Healthy) even both microenvironments have comparable vessel porosity. However, the presence of cancer cells such as MDA-MB-231 and C3Asub28 caused higher vessel permeability compared to endothelial monoculture (Control –, Tumorigenic) and THLE-3/endothelial co-culture, which agrees with previously published work in which co-culture of cancer cells increases vessel permeability (Buchanan *et al*., 2014). We can clearly observe the difference between normal liver and hepatocellular carcinoma as evidenced by cancerous cells increasing permeability by 2.77- (p<0.001) and 2.35-fold (p<0.05) for 70 and 3 kDa particles, respectively. Former studies on vascularized tumor-endothelial microenvironments also showed similar findings, in which an increase in transport of macromolecules occurred due to inclusion of cancer cells (Butler *et al*., 1975, Fukumura & Jain, 2007, Jain *et al*., 2014). There are two underlying reasons for this difference between the two liver cell lines. First, simulated drugs (dextran) have been perfused through normal liver with higher wall shear stress to generate physiological transport. Secondly, due to the interaction between cancer and endothelial cells or tumorigenic protein release by cancer cells, the endothelial layer porosity increased, which was discussed in the previous section. This phenomenon is described as the EPR effect and is more dominant compared to high wall shear stress (Buchanan *et al*., 2014).

The increase of vessel permeability in the presence of tumor cells strengthens the likelihood of cell invasion and migration into the endothelial layer, as suggested by former studies on vascularized tumor microenvironments (Buchanan *et al*., 2014, Kebers *et al*., 1998, Zervantonakis *et al*., 2012). Cancer cells affect endothelial integrity, as is evident from the large pores shown in Figure 6. This more porous endothelial layer is associated with higher permeability, as presented in Figure 7d. The vessel permeability and porosity are indicative of the transport properties; however, the impact of the particle size on vessel regulation is also a key factor in particle delivery and accumulation that needs to be taken into consideration. For all microenvironments, 3 kDa dextran particles were more permeable than 70 kDa. Other *in vivo* drug testing studies have also shown that permeability is highly affected by the size of nanoparticles (Dreaden *et al*., 2012, Terentyuk *et al*., 2009, Venkatasubramanian *et al*., 2008). The relationship between hydrodynamic diameter and permeability coefficient can be explained using Stokes-Einstein Equation of diffusivity (Yuan *et al*., 1995). By definition, particle size is indirectly proportional to permeability, which is indicated with higher diffusivity of smaller particles. Therefore, more rapid diffusion was observed for 3 kDa particles compared to 70 kDa, which resulted in a higher permeability coefficient. Moreover, the presence of the endothelial layer around the vasculature acted as an extra resistant layer and yielded a reduction in permeability coefficient of dextran particles, as shown in Figure 6.

The validity of permeability measurements were assessed by comparing the fold changes between the same Control – (Tumorigenic) and Control + findings reported in the literature. The collagen-based vascularized breast cancer platform developed by Zervantonakis *et al.* determined the fold change between Control – (Tumorigenic) and Control + as $1.79 \pm 0.27$ (Zervantonakis *et al*., 2012), which is in a good agreement with the quantified value of $1.59 \pm 0.13$ in this study. Moreover, a permeability coefficient of 70 kDa dextran particles was reported as $2.58 \times 10^{-6} \pm 0.19 \times 10^{-6}$ cm/s in vascularized collagen-based tumor microenvironments under the same shear stress (Buchanan *et al*., 2014, Michna *et al*., 2018), which compares well with our permeability results ($2.68 \times 10^{-6} \pm 0.29 \times 10^{-6}$ cm/s).

**Intensity Profiles and Accumulation:** In addition to the permeability, intensity profiles of particle fluorescence within the vessel and ECM provides insight regarding the accumulation of each type of particle in the different tissue microenvironments. Figure 8 presents intensity profiles for two different particle sizes and five different microenvironments (Control – (Healthy and Tumorigenic), C3Asub28/TIME, MDA-MB-231/TIME, and THLE-3/TIME). For all these microenvironments, we observed a sharp change in the slope between the vessel and the ECM interface. This is due to the presence of the endothelium, which acts as a barrier to particle transport. However, this decay significantly changes for different particle sizes. The intensity rate over time for small particle sizes is more rapid compared to large particles. This trend was experienced for all microenvironments, as smaller particles were able to penetrate faster through endothelial pores compared to larger particles (Zervantonakis *et al.*, 2012). This was anticipated since 70 kDa is a heavier solute and possesses a lower diffusivity compared to the 3 kDa particles. Additionally, the increase in maximum intensity over time with smaller particle size at the center of the vessel was detected for tumor cell lines. According to results shown in Figure 8 at t = 120 min, the peak intensity ratio of small particles to large for the MDA-MB-231 and C3Asub28 microenvironments was 1.66- and 1.59-fold, respectively. This trend can be explained by both collagen-based vascularized 3D *in vitro* tumor microenvironments and modeling studies (Buchanan *et al.*, 2014, M. Kim *et al.*, 2013, Wu *et al.*, 2013) for two main reasons: 1) Advective transport through the vessel is more dominant than diffusive Brownian motion into the ECM (M. Kim *et al.*, 2013, Wu *et al.*, 2013), and 2) Particles are diffusing and then leaving the ECM, which causes accumulation around the vessel (Buchanan *et al.*, 2014, M. Kim *et al.*, 2013). Given the fact that vessel porosity in tumor microenvironments is significantly higher compared to healthy tissue microenvironments (as seen in Figure 7a), particles can rapidly penetrate into the ECM. Furthermore, particles are able to freely diffuse back from the ECM to the vessel since the leaky endothelial layer fails to trap particles inside the ECM. The magnitude of the liver carcinoma intensity profile is much higher than all other microenvironments. This was anticipated since ECM and vessel porosity of liver cancer was much higher than other microenvironments, as presented in Figures 7a and 7c. Moreover, the peak intensity for liver carcinoma after 2 hours of perfusion (Figure 8) increased by 2.77- and 4.48-fold (p<0.01) for 3 and 70 kDa, respectively, compared to Control – (Tumorigenic). A similar trend was found for breast carcinoma, in which 1.39- and 3.65-fold (p<0.05) increases occurred for 3 and 70 kDa, respectively. On the other hand, healthy liver peak intensity did not change significantly compared to Control –, Healthy (Figure 8). In the context of breast carcinoma, we did not observe a substantial change between 3 kDa and 70 kDa particle diffusion (although this was observed for liver carcinoma). A possible explanation for this difference is that parameters other than vessel porosity, such as vessel pore structure and pore size, have contributed to nanoparticle transport.

Connecting the vascularized liver and breast tumor microenvironments in series and perfusing particles in either vessel enables simulation of the accumulation behavior of metabolization of particles (liver to tumor) or direct delivery to the tumor (tumor to liver). With both microenvironments connected, independent of which microenvironment received particles first, we noticed a substantial decrease in the magnitude of the intensity in the second microenvironment (Figure 9) for circulation in two microenvironments compared to perfusion through a single microenvironment alone (Figure 8). This would be the expected result due to the fact that the first microenvironment retains some portion of supplied particles. Moreover, the peak intensity value for MDA-MB-231/TIME microenvironments after passing through THLE-3/TIME microenvironments decreased by 2.40- and 1.99-fold ($p < 0.05$) for 3 and 70 kDa, respectively, compared to circulation of particle in the MDA-MB-231/TIME microenvironment alone. The likely explanation for this finding is that particles had already been uptaken by healthy liver cells, which simulates the drug being metabolized by the liver. Results from a clinical study suggest that the high accumulation of many chemotherapeutics in liver tissue results in injuries or liver failure (King & Perry, 2001). Our results show that large particles accumulate more than small particles in tumor cells. This outcome is also supported by an *in vivo* study, which found a 1.11-fold higher delivery for nanoparticles conjugated with chemotherapeutics, when compared to free chemotherapeutic drugs (Petryk *et al.*, 2013). However, when the healthy liver microenvironment was replaced with the liver tumor microenvironment, the downstream breast tumor microenvironment peak intensity increased by 1.31-fold ($p < 0.05$) for 3 kDa and decreased by 2.60-fold ($p < 0.05$) for 70 kDa. This interesting result implies that small particles were not trapped inside liver carcinoma platforms due to the leaky endothelial barrier but larger particles were trapped. This suggests further investigation if particles are actually accumulating in the ECM region or transporting back to the vessel. To gain further understanding, the intensity rates were determined using data from Figures 8 and 9. These calculated accumulation results in the vessel and ECM are presented in Figure 10. Intensity rates illustrated in Figure 10a show that the particle accumulation rate in the ECM in carcinoma microenvironments is significantly higher than in the healthy liver microenvironment. There are two possible explanations for this finding: i) the leakiness of the endothelial layer causes the EPR effect, and ii) significantly lower shear stress allows more time for particles to diffuse through the ECM (similar to work described by Buchanan *et al.* regarding permeability change with respect to wall shear stress (Buchanan *et al.*, 2014)).

Although we demonstrated that the ECM porosity of the THLE-3/TIME (healthy liver) microenvironment is higher than the ECM porosity of both cancer microenvironments, the particle accumulation rate in the MDA-MB-231/TIME and C3Asub28/TIME microenvironments are 3.45- and 4.81-fold ($p < 0.05$) higher than the healthy liver microenvironment, respectively. Similarly, a recent clinical study on drug delivery indicated that uptake by tumorigenic portions of liver is significantly higher than healthy portions (Haste *et al.*, 2017). This indicates that vessel porosity plays a more dominant role compared to ECM porosity in the accumulation rates of particles in the ECM. Vessel accumulation in the tumor microenvironments will enhance the likelihood of particles being delivered to other healthy tissues, causing toxicity (as is shown by previously published *in vivo* drug distribution studies (NDong *et al*., 2015, Petryk *et al*., 2013)). A higher ECM accumulation rate in the liver cancer compared to the breast cancer condition would be anticipated, based on the higher ECM porosity of liver cancer (delineated in Figure 7c). Moreover, Figure 10b shows that vessel accumulation rates for MDA-MB-231/TIME and C3Asub28/TIME microenvironments are 3.45- ($p<0.05$) and 8.11- ($p<0.01$) fold higher with smaller particle sizes compared to large ones. However, the vessel accumulation rate for large particle size was not significant except for MDA-MB-231/TIME microenvironments, with 17.67-fold change ($p<0.05$). This could be due to particles passing the leaky endothelial barrier and diffusing back to the vessel, which we do not observe for the healthy liver microenvironment because of the tight endothelial lumen (Chauhan *et al*. 2012). However, we did not observe similar phenomena for the C3Asub28/TIME microenvironment, even with a more porous endothelial layer. The endothelial structure in Figure 6 shows that MDA-MB-231 has a more patchy structure with more confluence, but in the C3Asub28 vessel we can see individual endothelial cells more distinctly. This patchy structure in MDA-MB-231 may be enabling the particles to be retained in the ECM but the frequent and individual gaps between endothelial cells in C3Asub28 may be allowing the particles to diffuse back to the vessel. A similar trend was described by a paper reporting that large and heterogeneous pores do not always retain nanoparticles, instead allowing them to diffuse back to the vessel for 1 nm and 12 nm nanoparticles (Chauhan *et al.* 2012), which are equivalent sizes to the dextran particles we have used. So potentially, the C3Asub28 vessel is within a threshold that allows molecules to not be retained and instead to move back out to the vessel, as opposed to MDA-MB-231. This suggests that the structure of the porosity is another deterministic factor on transport rates, which requires further investigation.

When multiple microenvironments are connected, the particle accumulation rate is expected to change due to particles remaining in the first-perfused microenvironment before entering the second microenvironment (Ma *et al*., 2012). Figure 10c presents the particle accumulation rate in the ECM for the four different microenvironment perfusion sequences considered in this study. Interestingly, the observed ECM accumulation rates were quite different between the THLE-3/TIME to MDA-MB-231/TIME perfusion sequence condition, and the converse MDA-MB-231/TIME to THLE-3/TIME perfusion sequence condition. For the small particle size, the ECM accumulation rate of MDA-MB-231/TIME decreased by 6.98-fold ($p<0.01$) after passing the THLE-3/TIME microenvironment, and the THLE-3/TIME ECM accumulation rate increased by 2.46-fold ($p<0.05$) after passing through the MDA-MB-231/TIME microenvironment. This result demonstrates the localization of free (unconjugated) chemotherapeutics within the liver for both metabolized and non-metabolized (direct delivery to tumor) conditions, which is disadvantageous for localizing chemotherapeutics in the tumor. When particles flowed from healthy liver to breast carcinoma platform, using large particles decreased the THLE-3/TIME ECM accumulation rate by 2.57-fold (p<0.01) and increased MDA-MB-231/TIME microenvironment accumulation rate by 5.57-fold (p<0.01) compared to small nanoparticles, as shown in Figure 10c. *In vivo* studies with similar particle sizes also found that nanoparticles with hydrodynamic diameter close to 15 nm have a greater accumulation rate in the tumor (Dreaden *et al.*, 2012, Terentyuk *et al.*, 2009). Moreover, vessel accumulation rates in multi microenvironment perfusion sequences were significantly decreased in the second microenvironment in all cases studied, as shown in Figure 10d. Based on these results, using chemotherapy alone may be less advantageous compared to chemotherapy-nanoparticle conjugation within the size range tested in this study, which agrees with findings from *in vivo* studies (NDong *et al.*, 2015, Petryk *et al.*, 2013). This outcome was observed for both cases in which we simulated the particles being metabolized (liver to tumor) and being directly delivered to the tumor (tumor to liver). When small particles were perfused through the healthy liver microenvironment first, the breast tumor microenvironment ECM accumulation rate was decreased by 5.49-fold (p<0.01) compared to perfusing through the tumor first, as presented in Figure 10d. In the simulated metabolized and direct-delivery cases where healthy liver was replaced with liver tumor cells, the particle accumulation rate was decreased by 1.05- and 3.94-fold (p<0.05) for 3 and 70 kDa particle sizes, respectively, as shown in Figure 10d. An overall summary of important accumulation results are summarized in Table 1.

**Overview of Results:** Small and large particles were selected to simulate chemotherapeutics and typical nanoparticle-conjugated chemotherapeutics, respectively. The use of large particles provides enhanced localization in the tumor site compared to smaller particles, which is due to leakiness of the tumor vasculature. Therefore, tumor blood vessels' porosity should be a consideration when selecting drug size to improve targeted drug delivery. Our findings are in agreement with to previous *in vivo* studies, which suggest that using chemotherapy drugs (without nanoparticle conjugation) at the size utilized in this study (~1 nm diameter) may not ensure derivation of the benefit of the EPR effect and targeted delivery (Blanco *et al.*, 2015, Jiang *et al.*, 2017, King & Perry, 2001). Moreover, biodistribution studies have also acknowledged that a particle diameter smaller than 5 nm is filtered by the kidney based on clinical and *in vivo* findings (Blanco *et al.*, 2015, Longmire *et al.*, 2008).

**Limitations and Future Works:** This study provides important findings for current passive transport investigations with respect to hydrodynamic diameter. However, hydrodynamic diameter is not the only aspect to consider in chemotherapy and chemotherapy-nanoparticle conjugation comparison. Different geometries of nanoparticles, including cylindrical and discoidal shapes can alter biodistribution and delivery characteristics such as circulation time, membrane interactions and macrophage uptake, among different organs (Blanco *et al*., 2015). Surface charge of the nanoparticle also provides distinct circulation lifetimes in different organs. For instance, Cabral *et al*. showed that negatively charged polymer micelle surfaces accumulate less in kidney and liver, which are the organs that drugs mainly accommodate (Cabral *et al.* 2011). Another method used to increase tumor uptake is coating nanoparticles. The widely used coating agent polyethylene glycol (PEG) has been shown to decrease nanoparticle aggregation, improve delivery to the tumor site, increment circulation time, limit Renal excretion created by the kidney, and diminish macrophage uptake by allowing additional anti-fouling properties (Blanco *et al.* 2015, Wilhelm *et al*. 2016). Despite its advantages, NDong *et al*. suggests that PEGylation can increase accumulation *in vivo* liver at certain nanoparticle sizes (NDong *et al*. 2015). Therefore, the impact of drug and nanoparticle properties on tumor uptake should be analyzed with alternative drug testing tools. Although our main interest was to investigate uptake differences of two identical dextran particles with varying particle size to replicate commonly used chemotherapeutic drugs and their conjugation with nanoparticles, the above-mentioned parameters are important future considerations.

**Conclusion**

This paper shows that the tumor and liver tissue-on-a-chip devices utilized in this study allow the high-throughput measurement of spatiotemporal biodistribution of therapeutic agents and nanoparticles while ensuring that their treatment responses are captured. By altering the direction of flow, we can simulate the effect of local delivery or metabolism by the liver on the transport kinetics of drugs and nanoparticles. The fabrication method for this platform is robust and flexible and can be adapted to mimic other tissue microenvironments. The developed tumor and liver-on-a-chip microenvironments can also be utilized for testing a combination of different treatment methods, such as hyperthermia, radiation, and a myriad of nanoparticles with unique functionality to create solutions for targeted delivery. Results from our experiments with these microenvironments are in agreement with results from comparable *in vivo* studies. Overall, tissue-on-a-chip devices inevitably have greater potential than standard cell culture, static *in vitro* setups. If the device is complex enough, it can augment or replace animal testing for advanced drug development before clinical studies.

**Acknowledgment**: We would like to acknowledge our funding from National Institute of Health Grant R21EB019646.

**Figure Legend**

**Figure 1:** Fabrication steps for the vascularized tissue microenvironment. PDMS was mixed with curing agent and poured into the aluminum mold shown in (I) and baked. Inlet and outlets were patterned around a 22G or 27G needle, and housing was patterned around the aluminum extrusion shown in (II). PDMS was peeled off from the aluminum mold and bonded to the glass slide and platform shown in (III), and was treated with PEI, glutaraldehyde and DI $H_2O$. The same platform cavity was filled with collagen mixture with the appropriate cell line. To form the channel to simulate the vessel, the needle was inserted (IV). Needle sizes were selected depending on desired wall shear stress. The needle was removed after polymerization of collagen (V) and preconditioned after injection of endothelium cells for 72 hours to form a vessel (VI). 3D illustration of the fabrication process is provided in the Supplement II.

**Figure 2:** Design and fabrication of the multi-chamber microfluidic platform and perfusion setup. (a) CAD design of the aluminum mold with 22G inlets. (b) Schematic of healthy liver-breast tumor microenvironment interaction and transport. (c) Closeup view of the platform with 0.5" 22G pins inserted into the chamber inlet and outlet for flow preconditioning and particle testing. Confocal images show preconditioned tumorigenic and healthy vessels with GFP-tagged breast cancer cells and FITC-tagged anti-Albumin immunostained healthy liver cells. Scale bar is 500 µm. (d) Shear stress profile across tumor and healthy vessels obtained using finite element method simulations. Unit of color gradient legend is dyn/cm$^2$. Targeted physiological wall shear stresses for tumorigenic and healthy vessels are 1 and 4 dyn/cm$^2$, respectively; consequently, vessel diameters for tumorigenic and healthy platforms are 711 and 435 µm.

**Figure 3:** Morphology of C3Asub28, MDA-MB-231, and THLE-3 cell lines within the avascular microenvironments. F-Actin and DAPI stained samples show aggregation over time. Scale bar is 20 µm. SEM images show the outline of a single cell in each matrix on day 3. Scale bar is 10 µm.

**Figure 4**: CellTiter-Blue viability assay results show cells growth over time within the avascular platforms, with initial seeding density of 1x106 cells/mL for THLE-3, MDA-MB-231, and C3Asub28 cell lines within the tissue microenvironments over 3 days. Cell concentration was normalized to Day 0. Statistical significance was compared to Day 0. Data shown are mean ± SD. (n=5, ∗ p<0.05).

**Figure 5:** Albumin expression and release from healthy liver cells within THLE-3/TIME vascularized microenvironments. a) FITC-tagged anti-albumin immunostained healthy liver cells overlaid with bright field image. Scale bar is 10 µm, b) Albumin release from THLE-3/TIME vascularized microenvironment during days

one, two, and three (the preconditioning period), and then under physiological wall shear stress after the preconditioning period. Statistical significance was compared to Day 1. Data shown are mean $\pm$ SD (n = 5, ∗∗∗: p<0.005, ∗∗∗∗: p<0.001).

**Figure 6:** Confocal images of mKate-labeled (red) endothelial cells in each vascularized in vitro microenvironment. Control – vessels refer to TIME monoculture under conditions corresponding to the two different collagen concentrations and vessels sizes and wall shear stresses used in this study (Healthy and Tumorigenic), and Control + refers to TIME monoculture in Tumorigenic-condition collagen concentration/vessel size treated with TNF-α. Vasculature diameter varies between 411-450 (for THLE-3 and Control –, Healthy) and 700-750 µm (for all other cell lines). Control - (Healthy) and THLE-3 vessels are exposed to 4 dyn/cm$^2$ wall shear stress and all other vessel are exposed to 1 dyn/cm$^2$ wall shear stress. Scale bar is 500 µm.

**Figure 7:** Permeability and porosity for different cell culture microenvironments. a) Quantified vessel porosities using confocal microscopy images, b) Fiber structure of different microenvironments obtained with SEM images. Scale bar is 5 µm, c) Quantified ECM porosity of each microenvironment using SEM images, d) Permeability of endothelial lumen for different particle sizes and vascularized microenvironments. Control - (Tumorigenic) refers to TIME monoculture with the collagen concentration of 7 mg/ml, 711 µm vessel diameter and exposed to 1 dyn/cm$^2$ wall shear stress. Control - (Healthy) refers to TIME monoculture with the collagen concentration of 4 mg/ml, 435 µm vessel diameter and exposed to 4 dyn/cm$^2$ wall shear stress. Control + refers to TNF-α treated Control – (Tumorigenic) vasculature. Data shown are mean $\pm$ SD (n = 4, ∗: p<0.05, ∗∗: p<0.01, ∗∗∗: p<0.005, ∗∗∗∗: p<0.001, n.s.: not significant).

**Figure 8:** Diffusion curves of fluorescent dextran particles (3 and 70 kDa) with respect to time and position across ECM boundaries for healthy liver (THLE-3), and tumorigenic breast (MDA-MB-231) and liver (C3Asub28) microenvironments and their corresponding controls. Control - (Tumorigenic) refers to TIME monoculture with the collagen concentration of 7 mg/ml, 711 µm vessel diameter and exposed to 1 dyn/cm$^2$ wall shear stress. Control - (Healthy) refers to TIME monoculture with the collagen concentration of 4 mg/ml, 435 µm vessel diameter and exposed to 4 dyn/cm$^2$ wall shear stress. Dashed lines represent vessel boundaries. Fluorescence intensity profiles of three experiments were averaged (n=3).

**Figure 9:** Diffusion curves of fluorescent dextran particles (3 and 70 kDa) with respect to time and position across ECM boundaries for secondary compartments of tumorigenic breast (MDA-MB-231) and liver (C3Asub28), and healthy liver (THLE-3) microenvironments. Control - (Tumorigenic) refers to TIME monoculture with the collagen concentration of 7 mg/ml, 711 µm vessel diameter and exposed to 1 dyn/cm$^2$ wall shear stress. Control -

(Healthy) refers to TIME monoculture with the collagen concentration of 4 mg/ml, 435 μm vessel diameter and exposed to 4 dyn/cm$^2$ wall shear stress. Dashed lines represent vessel boundaries. The microenvironment sequence order of nanoparticle circulation is presented as in the figure. Fluorescence intensity profiles of three experiments were averaged (n=3).

**Figure 10:** Accumulation rate in vessel and ECM in response to different sequences of perfusion in the tissue microenvironments. First microenvironment accumulation rates in the a) ECM and b) vessel for THLE-3, MDA-MB-231, and C3Asub28 and their corresponding control microenvironments. Second microenvironment accumulation rate in the c) ECM and d) vessel for MDA-MB-231, C3Asub28, and THLE-3 microenvironment interactions when particles are introduced to either the healthy or tumorigenic platforms first. Microenvironment perfusion sequence order was applied as described in the figure. Control - (Tumorigenic) refers to TIME monoculture with the collagen concentration of 7 mg/ml, 711 μm vessel diameter and exposed to 1 dyn/cm$^2$ wall shear stress. Control - (Healthy) refers to TIME monoculture with the collagen concentration of 4 mg/ml, 435 μm vessel diameter and exposed to 4 dyn/cm$^2$ wall shear stress. Data shown are mean ± SD. (n = 4, *: p<0.05, **: p<0.01, ***: p<0.005, ****: p<0.001, n.s.: not significant.)

**Table Legend**

**Table 1:** Summary of important transport findings for different platforms.

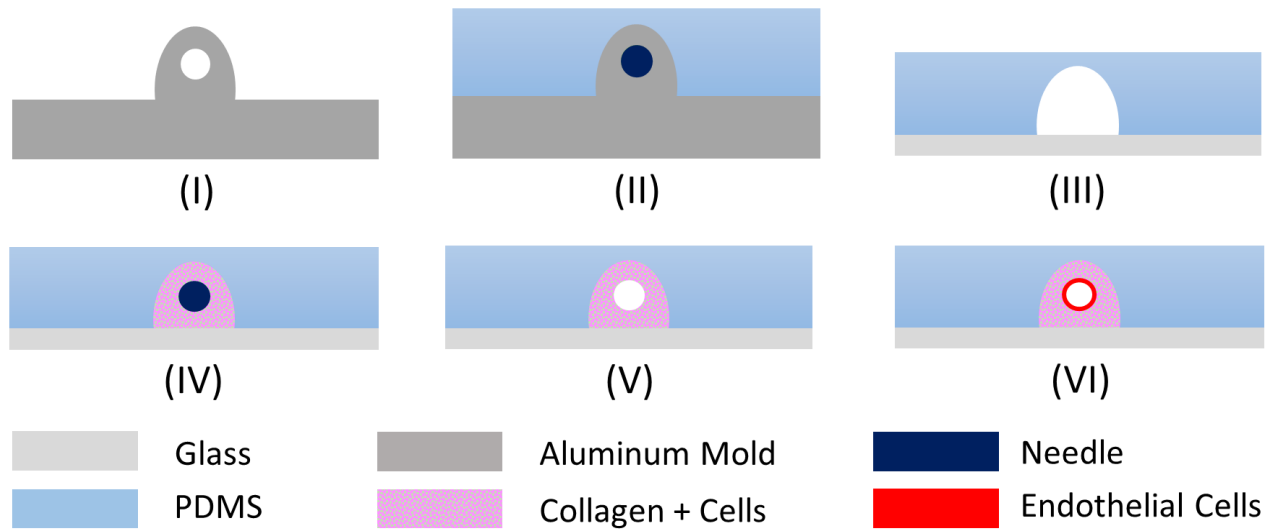

**Figure 1:** Fabrication steps for the vascularized tissue microenvironment. PDMS was mixed with curing agent and poured into the aluminum mold shown in (I) and baked. Inlet and outlets were patterned around a 22G or 27G needle, and housing was patterned around the aluminum extrusion shown in (II). PDMS was peeled off from the aluminum mold and bonded to the glass slide and platform shown in (III), and was treated with PEI, glutaraldehyde and DI $H_2O$. The same platform cavity was filled with collagen mixture with the appropriate cell line. To form the channel to simulate the vessel, the needle was inserted (IV). Needle sizes were selected depending on desired wall shear stress. The needle was removed after polymerization of collagen (V) and preconditioned after injection of endothelium cells for 72 hours to form a vessel (VI). 3D illustration of the fabrication process is provided in the Supplement II.

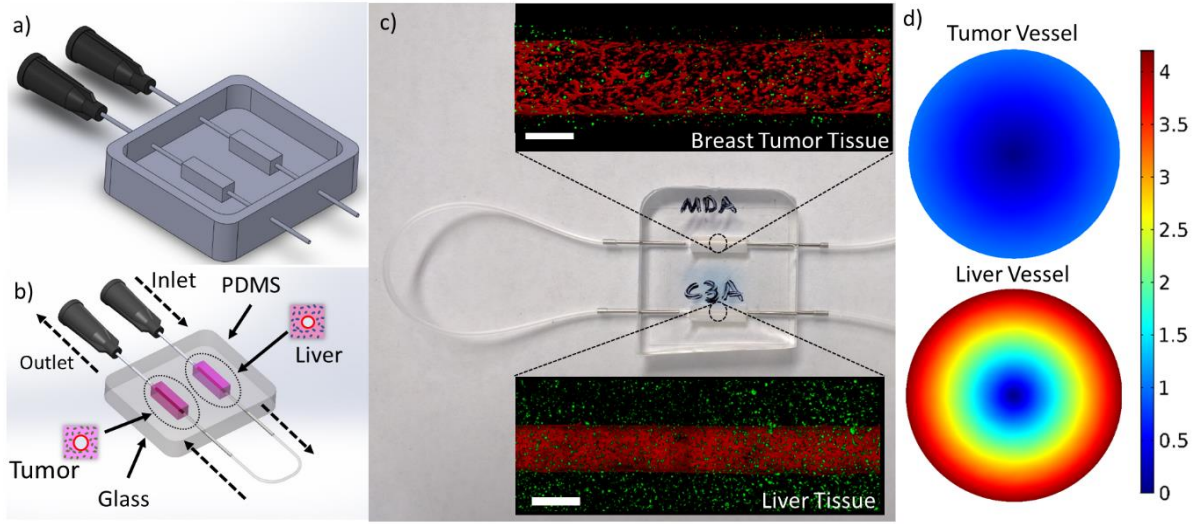

**Figure 2:** Design and fabrication of the multi-chamber microfluidic platform and perfusion setup. (a) CAD design of the aluminum mold with 22G inlets. (b) Schematic of healthy liver-breast tumor microenvironment interaction and transport. (c) Closeup view of the platform with 0.5" 22G pins inserted into the chamber inlet and outlet for flow preconditioning and particle testing. Confocal images show preconditioned tumorigenic and healthy vessels with GFP-tagged breast cancer cells and FITC-tagged anti-Albumin immunostained healthy liver cells. Scale bar is 500 µm. (d) Shear stress profile across tumor and healthy vessels obtained using finite element method simulations. Unit of color gradient legend is dyn/cm$^2$. Targeted physiological wall shear stresses for tumorigenic and healthy vessels are 1 and 4 dyn/cm$^2$, respectively; consequently, vessel diameters for tumorigenic and healthy platforms are 711 and 435 µm.

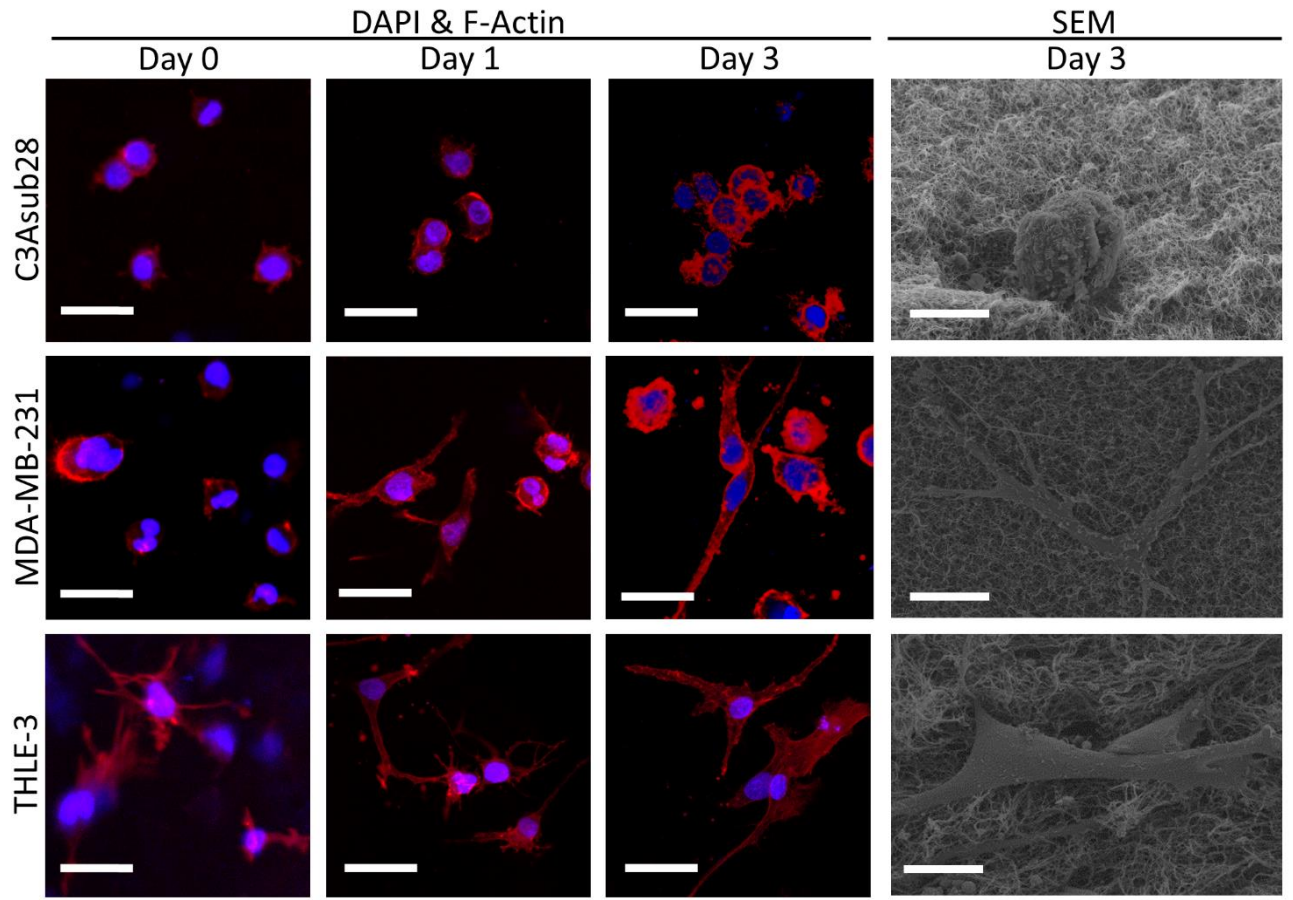

**Figure 3:** Morphology of C3Asub28, MDA-MB-231, and THLE-3 cell lines within the avascular microenvironments. F-Actin and DAPI stained samples show aggregation over time. Scale bar is 20 µm. SEM images show the outline of a single cell in each matrix on day 3. Scale bar is 10 µm.

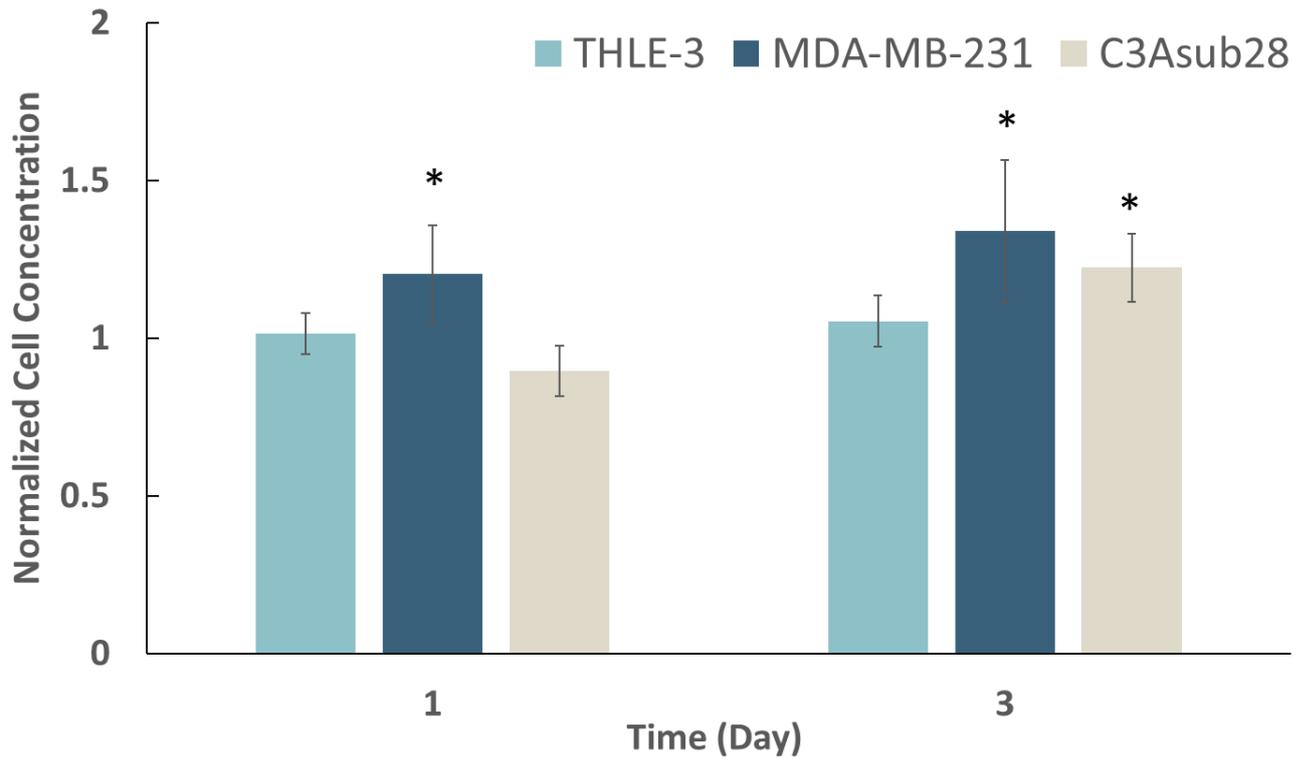

**Figure 4:** CellTiter-Blue viability assay results show cells growth over time within the avascular platforms, with initial seeding density of $1x10^6$ cells/mL for THLE-3, MDA-MB-231, and C3Asub28 cell lines within the tissue microenvironments over 3 days. Cell concentration was normalized to Day 0. <span style="color:red">Statistical significance was compared to Day 0.</span> Data shown are mean ± SD. (n=5, ∗ p<0.05).

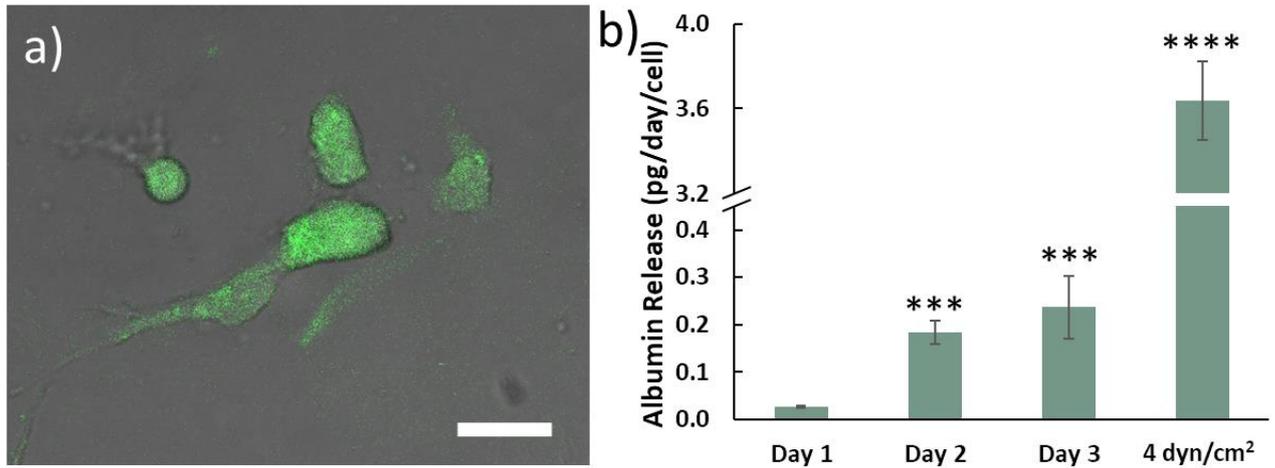

**Figure 5:** Albumin expression and release from healthy liver cells within THLE-3/TIME vascularized microenvironments. a) FITC-tagged anti-albumin immunostained healthy liver cells overlaid with bright field image. Scale bar is 10 µm, b) Albumin release from THLE-3/TIME vascularized microenvironment during days one, two, and three (the preconditioning period), and then under physiological wall shear stress after the preconditioning period. <span style="color:red">Statistical significance was compared to Day 1.</span> Data shown are mean ± SD (n = 5, ***: p<0.005, ****: p<0.001).

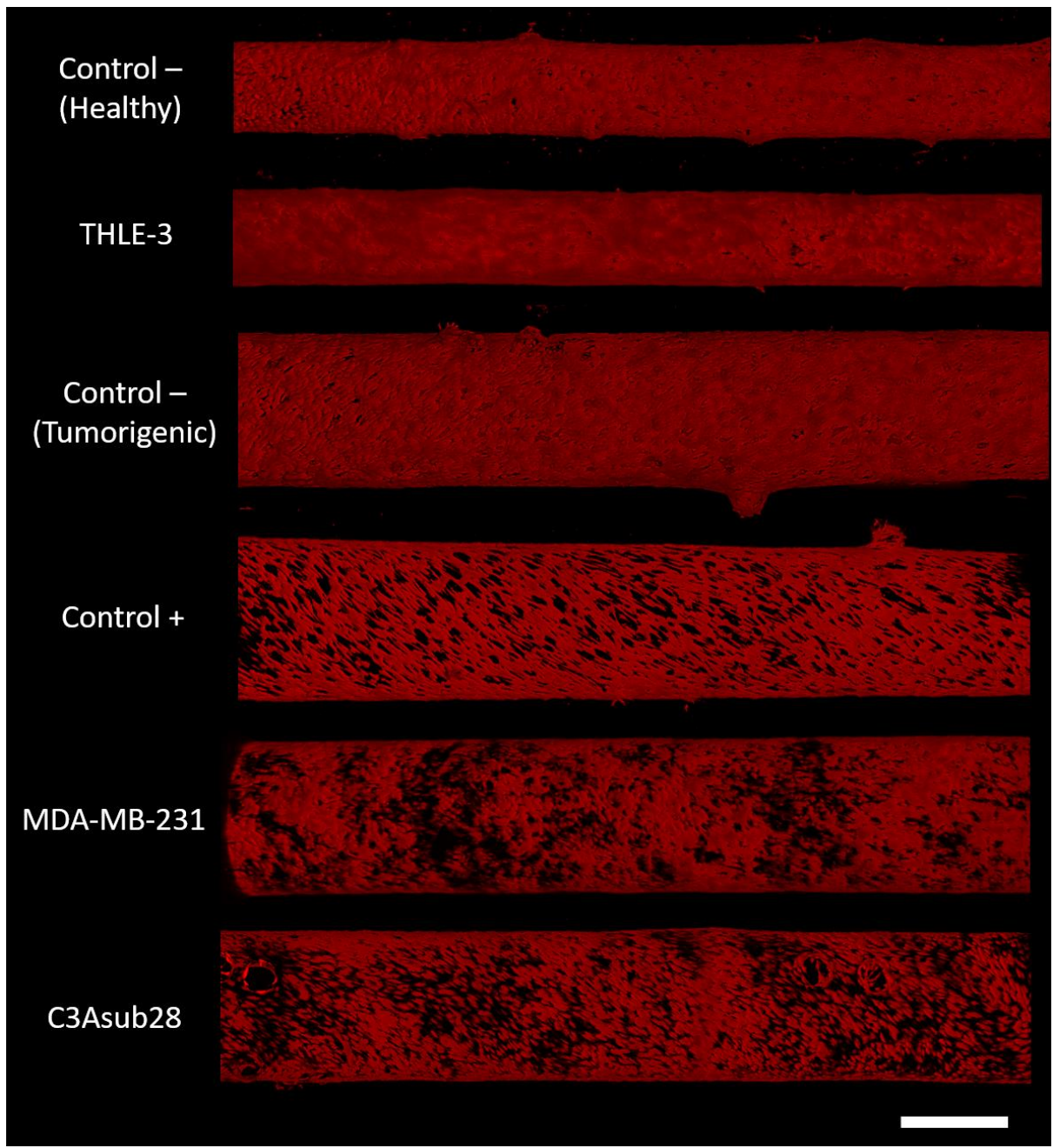

**Figure 6:** Confocal images of mKate-labeled (red) endothelial cells in each vascularized *in vitro* microenvironment. Control – vessels refer to TIME monoculture under conditions corresponding to the two different collagen concentrations and vessels sizes and wall shear stresses used in this study (Healthy and Tumorigenic), and Control + refers to TIME monoculture in Tumorigenic-condition collagen concentration/vessel size treated with TNF-α. Vasculature diameter varies between 411-450 (for THLE-3 and Control –, Healthy) and 700-750 µm (for all other cell lines). Control - (Healthy) and THLE-3 vessels are exposed to 4 dyn/cm$^2$ wall shear stress and all other vessel are exposed to 1 dyn/cm$^2$ wall shear stress. Scale bar is 500 µm.

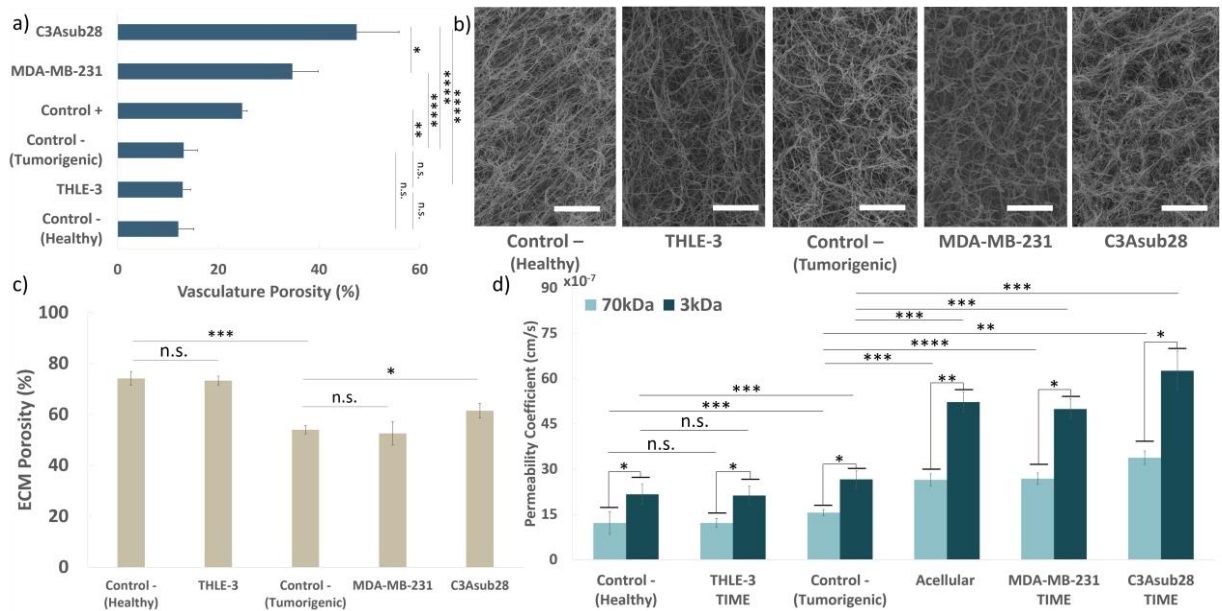

**Figure 7:** Permeability and porosity for different cell culture microenvironments. a) Quantified vessel porosities using confocal microscopy images, b) Fiber structure of different microenvironments obtained with SEM images. Scale bar is 5 µm, c) Quantified ECM porosity of each microenvironment using SEM images, d) Permeability of endothelial lumen for different particle sizes and vascularized microenvironments. Control - (Tumorigenic) refers to TIME monoculture with the collagen concentration of 7 mg/ml, 711 µm vessel diameter and exposed to 1 dyn/cm$^2$ wall shear stress. Control - (Healthy) refers to TIME monoculture with the collagen concentration of 4 mg/ml, 435 µm vessel diameter and exposed to 4 dyn/cm$^2$ wall shear stress. Control + refers to TNF-$\alpha$ treated Control – (Tumorigenic) vasculature. Data shown are mean ± SD (n = 4, *: p<0.05, **: p<0.01, ***: p<0.005, ****: p<0.001, n.s.: not significant).

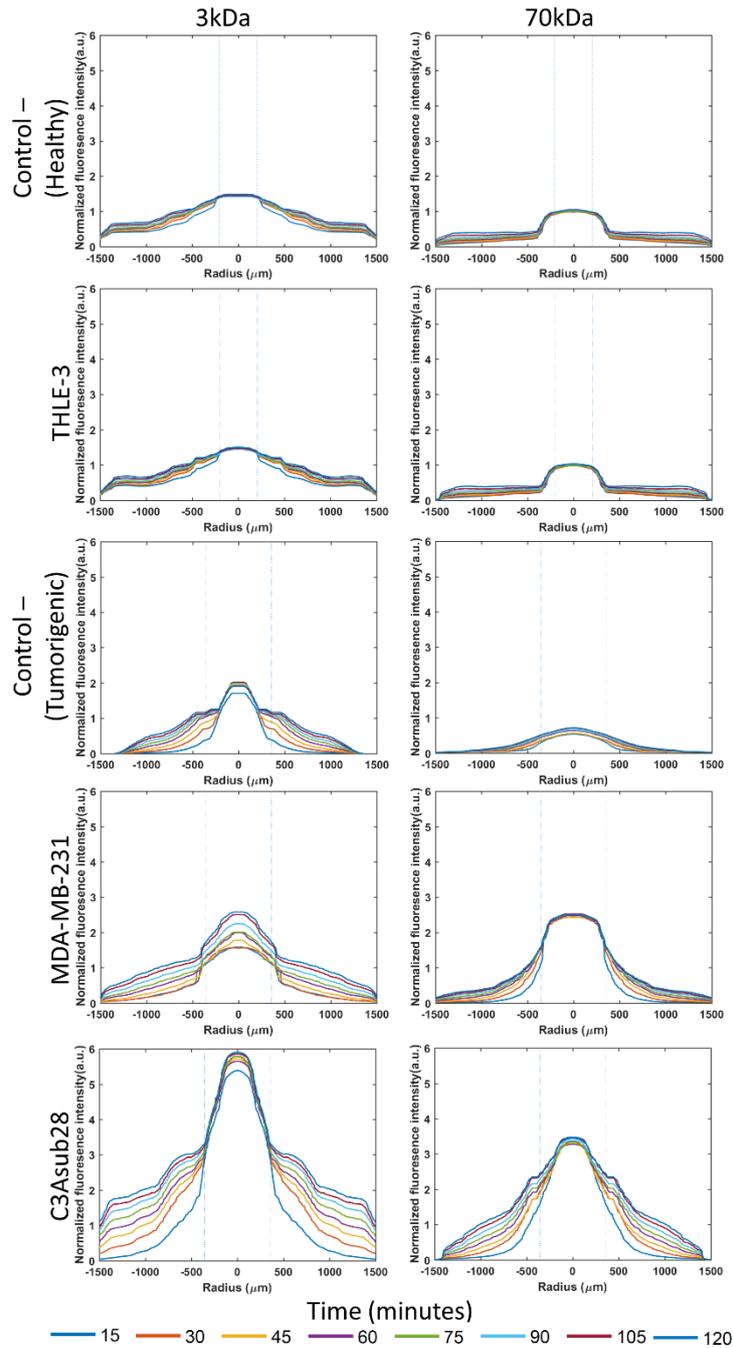

**Figure 8:** Diffusion curves of fluorescent dextran particles (3 and 70 kDa) with respect to time and position across ECM boundaries for healthy liver (THLE-3), and tumorigenic breast (MDA-MB-231) and liver (C3Asub28) microenvironments and their corresponding controls. Control - (Tumorigenic) refers to TIME monoculture with the collagen concentration of 7 mg/ml, 711 µm vessel diameter and exposed to 1 dyn/cm$^2$ wall shear stress. Control - (Healthy) refers to TIME monoculture with the collagen concentration of 4 mg/ml, 435 µm vessel diameter and exposed to 4 dyn/cm$^2$ wall shear stress. Dashed lines represent vessel boundaries. Fluorescence intensity profiles of three experiments were averaged (n=3).

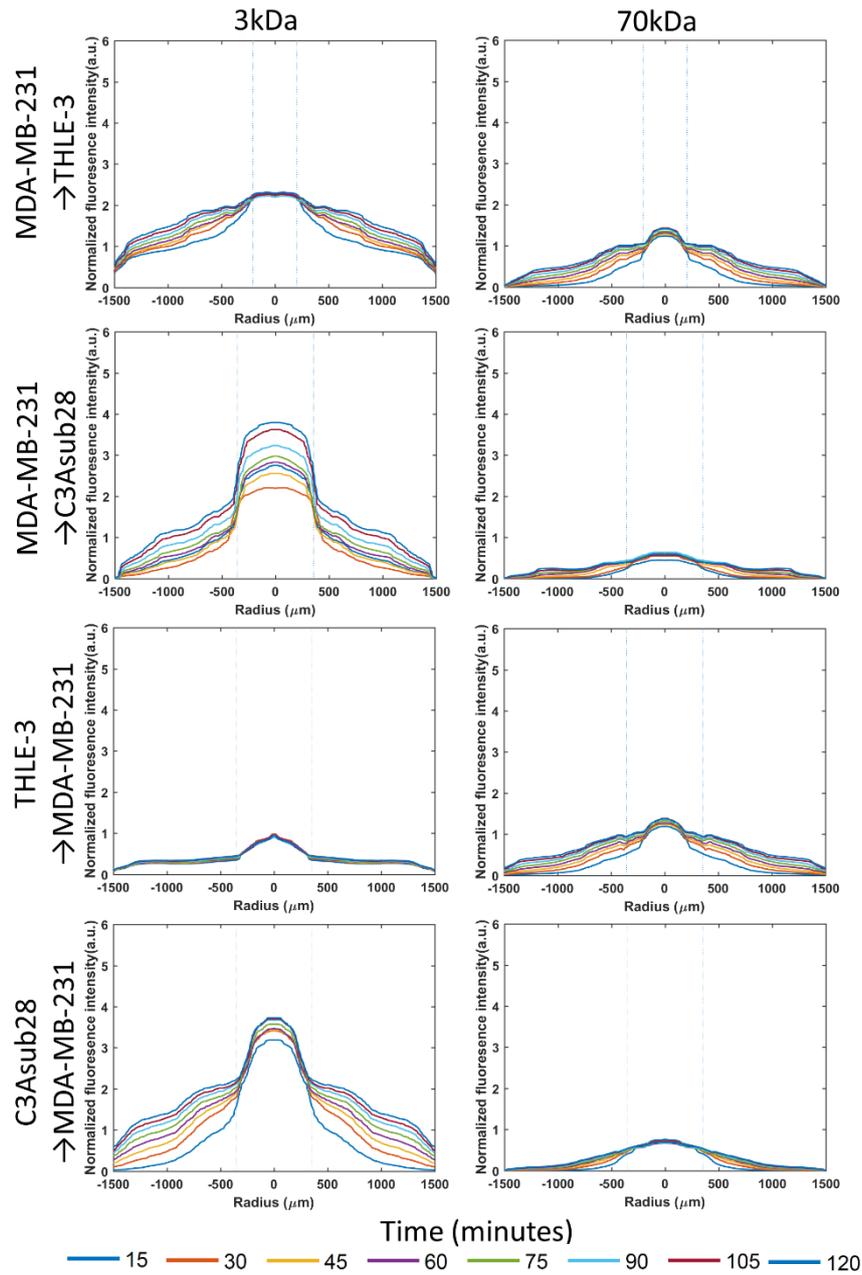

**Figure 9:** Diffusion curves of fluorescent dextran particles (3 and 70 kDa) with respect to time and position across ECM boundaries for secondary compartments of tumorigenic breast (MDA-MB-231) and liver (C3Asub28), and healthy liver (THLE-3) microenvironments. Control - (Tumorigenic) refers to TIME monoculture with the collagen concentration of 7 mg/ml, 711 µm vessel diameter and exposed to 1 dyn/cm² wall shear stress. Control - (Healthy) refers to TIME monoculture with the collagen concentration of 4 mg/ml, 435 µm vessel diameter and exposed to 4 dyn/cm² wall shear stress. Dashed lines represent vessel boundaries. The microenvironment sequence order of nanoparticle circulation is presented as in the figure. Fluorescence intensity profiles of three experiments were averaged (n=3).

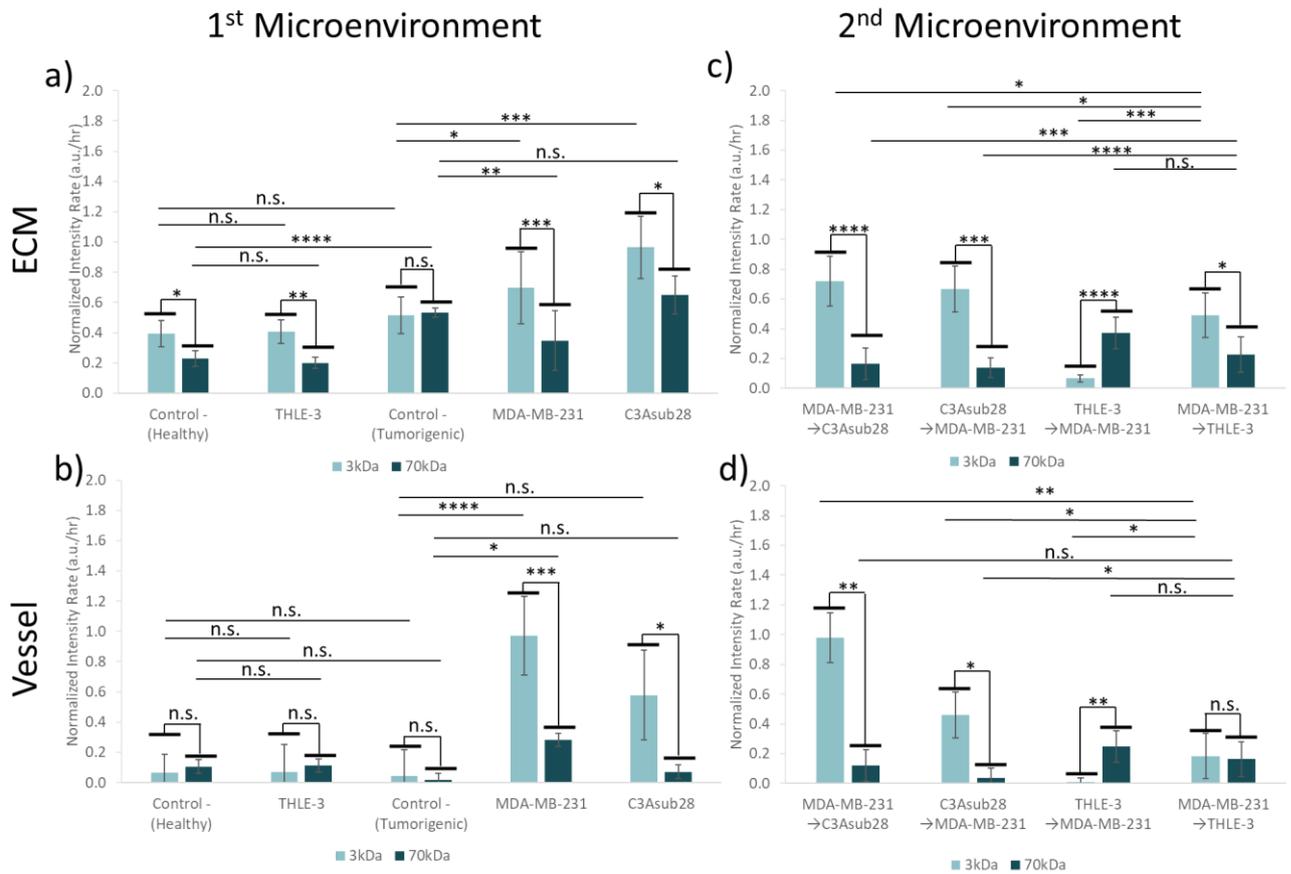

**Figure 10:** Accumulation rate in vessel and ECM in response to different sequences of perfusion in the tissue microenvironments. First microenvironment accumulation rates in the a) ECM and b) vessel for THLE-3, MDA-MB-231, and C3Asub28 and their corresponding control microenvironments. Second microenvironment accumulation rate in the c) ECM and d) vessel for MDA-MB-231, C3Asub28, and THLE-3 microenvironment interactions when particles are introduced to either the healthy or tumorigenic platforms first. Microenvironment perfusion sequence order was applied as described in the figure. Control - (Tumorigenic) refers to TIME monoculture with the collagen concentration of 7 mg/ml, 711 μm vessel diameter and exposed to 1 dyn/cm² wall shear stress. Control - (Healthy) refers to TIME monoculture with the collagen concentration of 4 mg/ml, 435 μm vessel diameter and exposed to 4 dyn/cm² wall shear stress. Data shown are mean ± SD. (n = 4, *: p<0.05, **: p<0.01, ***: p<0.005, ****: p<0.001, n.s.: not significant.)

**Table 1:** Summary of important transport findings for different platforms.

| Cell Lines | Platform | Finding |
|---|---|---|
| MDA-MB-231/TIME | Breast Carcinoma | 70 kDa particle peak intensity was increased by 1.66-fold compared to 3 kDa particle. |
| | | 3 kDa and 70 kDa peak intensity was increased by 1.39- and 3.65-fold compared to Control - (Tumorigenic). |
| | | 70 kDa ECM accumulation rate increased by 5.57-fold compared to 3 kDa particles after passing through healthy liver. |
| | | For the metabolized condition, the 3 kDa vessel accumulation rate decreased by 5.49-fold compared to the direct delivery condition. |
| C3Asub28/TIME | Hepatocellular Carcinoma | 70 kDa particle peak intensity was increased by 1.59-fold compared to 3 kDa particle. |
| | | 3 kDa and 70 kDa peak intensity was increased by 2.77- and 4.88-fold compared to Control - (Tumorigenic). |
| | | 3 kDa and 70 kDa vessel accumulation increased by 1.05- and 3.94-fold respectively compared to the standalone condition after passing through breast carcinoma. |
| THLE-3/TIME | Healthy Liver | 70 kDa healthy liver ECM accumulation rate decreased by 2.57-fold compared to 3 kDa particles after passing through breast carcinoma. |
| | | 3 kDa ECM accumulation rate increased by 2.46-fold after particles passed through the breast carcinoma. |